\documentclass[proof]{WileyASNA-v1}

\articletype{Article Type}

\received{4 March 2023}
\revised{  2023}
\accepted{ 2023}

\begin{document}

\title{On the Chemical Composition of the Evolved Very Bright Metal-Poor Star HD~1936}


\author[1,2]{Seyma \c{C}al{\i}\c{s}kan}

\author[1]{Jannat Mushreq Kamil Alazzawi}

\author[1]{Yahya Nasolo}

\authormark{\c{C}al{\i}\c{s}kan \textsc{et al}}%

\address[1]{\orgdiv{Ankara University}, \orgname{Science Faculty, Department of Astronomy and Space Sciences}, \orgaddress{\state{Ankara}, \country{Turkey}}}
\address[2]{\orgdiv{Ankara University}, \orgname{Astronomy and Space Sciences Research and Application Center (Kreiken Observatory)}, \orgaddress{\state{Ankara}, \country{Turkey}}}

\corres{\c{S}eyma \c{C}al{\i}\c{s}kan, Ankara University, Science Faculty, Department of Astronomy and Space Sciences, Ankara, Turkey. ~ \email{seyma.caliskan@science.ankara.edu.tr}}


\abstract{We present chemical abundances of the very bright metal-poor star HD~1936 based on high-resolution and high SNR spectra from AUKR. We obtain the abundances of 29 atomic species with atomic numbers between 3 and 63. In this context, the derived lithium abundance of 1.01 is consistent with the thin Li plateau observed in lower red giant branch stars. The star is a carbon-normal with the ratio of -0.31, just like other low-luminosity red giants on the thin Li plateau. We find the ratios of [Eu/Fe]=0.43 and [Ba/Eu]=-0.64, indicating very little s-process contamination. These ratios allow us to classify the star as a moderately r-process-enhanced (r-I) metal-poor star for the first time. It is worth mentioning that the star has a metallicity of -1.74, a [Cu/Fe] of -0.74, a [Zn/Fe] of 0.04, and a [Mg/C] of 0.69. The results suggest that it may be a second-generation star formed in a multi-enriched environment, rather than being a descendant of very massive first-generation stars. A last point worth mentioning is the possibility that HD~1936 may host a sub-stellar component with a mass of 18.35$M_{\rm J}$. Although our study does not confirm or deny this, we briefly discuss the possibility of the star hosting a planet.}


\maketitle

\footnotetext{\textbf{Abbreviations:} IRAF, image reduction and analysis facility; NIST, National Institute of Standard and Technology; IDL, Interactive Data Language}

\section{Introduction}\label{sec1}

The hypothetical first stars in the universe, Pop~III, did not contain any metal, because no chemical enrichment had occurred in the environment in which they formed. A great deal of research has examined the existence of these stars and the mass limitations they might have if they exist. The most massive of these stars would have ended their lives with various kinds of supernova (SN) explosions (e.g., core-collapse SNe, hypernovae, pair-instability SNe) and polluted the environment with the metals they produced. The next generation of stars, second-generation Pop~II, formed in an environment polluted by the SNe of the first stars.

The chemical abundance pattern of a star is effectively its DNA, and chemical abundance analysis is necessary to understand its formation history. Metal-poor field stars are invaluable objects since they preserve the chemical properties of their progenitors, the first stars, in their own atmospheres. Thus, detailed atmospheric abundances for these stars provide an understanding of the chemical evolution of our Galaxy. 

The stars with intermediate-metallicity (-2.5<[Fe/H]<-1) could have been formed from gas polluted by the ejecta of pair-instability SNe, allowing us to probe the very massive first-generation of stars with masses ranging from 140 to 260$M_{\rm \odot}$~\citep{salvadorietal19}. If the intermediate-metallicity stars are indeed descendants of the very massive first-generation, one should expect to find an under-abundance of nitrogen, copper, and zinc in the sample star~\citep{salvadorietal19}. On the other hand, \citet{hartwigetal18} has suggested a mass range of 3 to 150$M_{\rm \odot}$ for the first stars, and focused on the fraction of second-generation stars formed in an environment (mono- and multi-) enriched by core-collapse or pair-instability SNe at [Fe/H]<-3. Their model results point against pair-instability SNe from stars with masses over 200$M_{\rm \odot}$. They also found that the all of second-generation stars with [Fe/H]$\leq$-7 are mono-enriched, along with about half those in the range -6<[Fe/H]<-4. While the majority of mono-enriched stars have [Fe/H]<$-$2, they can also exist at metallicities close to solar. Therefore, metallicity alone is not definitive indicator of whether a star is mono- or multi-enriched. The [Mg/C] ratio is more useful for determining the probability of mono-enrichment. A lower value of [Mg/C] may suggest the existence of mono-enriched second-generation stars~\citep{hartwigetal18}.

According to the table of nomenclature for stars of different metallicity published by~\citep{beers05}, HD~1936 is a metal-poor star with a metallicity of -1.73$\pm$0.08~\citep{pereiraetal19}. The star is located in the Galactic halo at a distance of 306.04~pc~\citep{dekaetal18}. HD~1936 (HIP~1873) is an interesting object because it has a physical component that is a sub-stellar object with a mass of 18.35$M_{\rm J}$, according to~\citet{kervellaetal19}. In their study, the binarities of nearby objects were determined from their proper motion anomalies, using Gaia DR2 and Hipparcos data. The mass and radius of the primary component of HD~1936 were found to be 2.92$M_{\rm \odot}$ and 12.12$R_{\rm \odot}$, respectively. They also provide a value of 2.14~AU for the orbital radius of the system, and -5.98~km/s for the heliocentric radial velocity. There are differing values for the radial velocity of the star in the literature.~\citet{brandt21} gives a radial velocity of -6.46$\pm$0.17~km/s from Gaia DR2 data.~\citet{niedzielskietal16} measured the radial velocity of the star as -7.73~km/s from ESO-FEROS data from 2006; the atmospheric and physical parameters of HD~1936 were determined in the same study. In~\citet{dekaetal18}, the mass, radius, luminosity, and age were re-calculated as 0.89$M_{\rm \odot}$, 9.93$R_{\rm \odot}$, 1.9$L_{\rm \odot}$, and 9.97~Gyr, and its kinematic properties were also examined. The atmospheric parameters of HD~1936 have also been determined by~\citep{pereiraetal19}, as well as its sodium abundance, but no detailed chemical abundance has been published for the star. Therefore, its chemical abundance is analyzed here for the first time.

Our work is aimed at performing the detailed abundance analysis of HD~1936 with intermediate-metallicity, using a small-size AUKR telescope. The detailed abundances of the star enable us to explore whether it is a descendant of pair-instability SN and which polluters contributed to the enrichment of its formation environment.

The paper is organized as follows. In Sec.~\ref{sec2}, we describe the observations and data reduction. Details on the stellar parameters are given in Sec.~\ref{sec3}. The chemical abundance analysis and our results are summarized in Sec.~\ref{sec4} and Sec.~\ref{sec5}, respectively. We discuss the evolutionary status of the star in Sec.~\ref{sec6}. We conclude in Sec.~\ref{sec7}.

\section{Observations and data reduction}\label{sec2}

\begin{figure}
\begin{center}
\hbox{\hspace{-0.2cm}\includegraphics[width=\columnwidth]{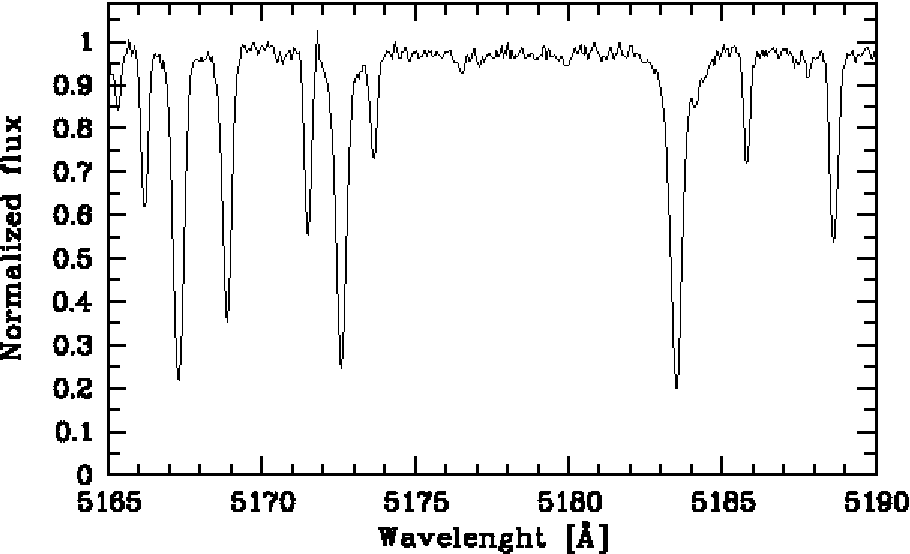}}
\caption{A sample of spectrum centered on the magnesium triplet.}
\label{mgtriplet}
\end{center}
\end{figure}

HD~1936 ($\alpha_{2000}=00^h23^m40^s$; $\delta_{2000}$=-03$^{\circ}$03' 29'') is a very bright metal-poor star of V$\sim$7.78 magnitude. It is located in the Pisces constellation. We obtained its optical region spectrum with the Whoppshel Spectrograph mounted at the 0.8-m Prof. Dr. Berahitdin Albayrak telescope of Ankara University Kreiken Observatory (AUKR). A high-resolution (R$\sim$30,000) and high signal-to-noise ratio (SNR$\sim$100 @5300\AA) spectrum of the star was taken with an exposure time of 3600~s, covering a wavelength range of 4000 to 7500~\AA~on November~12,~2022. Calibration images (bias, dark, flat field, Th-Ar lamp spectra) were taken in addition to the scientific images during this observation night. We used the AUKR data reduction pipeline for reducing our raw spectra. We then normalized the spectrum using standard IRAF packages~\citep{1986SPIE..627..733T,1993ASPC...52..173T}. A sample of the observed spectrum centered on the magnesium triplet region is shown in Fig.~\ref{mgtriplet}.  

By using unblended and moderately strong lines of titanium and iron, we calculated the heliocentric radial velocity of the star as -6.8$\pm$0.5~km/s and shifted the spectrum according to this value. The projected rotational velocity ($v$sin$i$) of the star was found to be 5$\pm3$~km/s, derived by comparing the observed and synthetic iron lines. We have noticed that the radial velocity of the star is not constant over the time. Systematic monitoring of the radial velocity of the star would be required to verify the possible binary nature of HD~1936. 

\section{Stellar Parameters}\label{sec3}
We first derived the effective temperature of HD~1936 from broad- (BVJHK) and narrow- (by) band optical photometry, using the calibrations of~\citet{alonsoetal99}. The photometric data were taken from~\citet{hogetal00}, \citet{cutrietal03}, and \citet{paunzen15}. Each color was corrected for the interstellar reddening~\citep{schlegeletal98}. The surface gravity of our sample star has been estimated by 10~Gyr Yale-Yonsei Isochrones~\citep{demarqueetal04} with the [Fe/H]=-1.5, [$\alpha$/Fe]=0.3, and photometrically derived T$_{\rm eff}$. 

We adopted an initial model atmosphere, including photometric $T_{\rm eff}$ and log~$g$ values, with the ATLAS9 code~\citep{kurucz93, sbordoneetal04}, assuming a plane-parallel geometry for line formation in local thermodynamic equilibrium (LTE) and with convection turned off. We then obtained the atmospheric parameters of the star using traditional spectroscopic methods: the excitation equilibrium of Fe~I lines for the effective temperature, a zero-slope equivalent width versus log $\epsilon$ plot of Fe~I lines for the microturbulence, and the ionization equilibrium of Fe~I/II lines for the surface gravity. To derive the atmospheric parameters, we employed the 91 Fe~I and 8 Fe~II lines. We then used the spectroscopic atmospheric parameters to perform the chemical abundance analysis. The photometrically and spectroscopically obtained parameters are given in Table~\ref{tab3}.

\begin{table*}
\caption{Atmospheric parameters of HD~1936.}
\label{tab3}
\centering
\begin{tabular*}{35pc}{lccccccccc}
\toprule
  & &  & T$_{\rm eff}$& & &log~$g$ & $\xi$& [Fe~I/H]&[Fe~II/H]\\
  & &  & (K)& &&(dex) & (km/s)  & & \\
\multicolumn{5}{c}{}& & & & \\
\midrule
&B-V&V-K&J-H&J-K&b-y& & & \\
\midrule
Photometric &4918 &4736 &5173 &5136 & 4833& & & \\
Photometric mean& & &4912 & & &2.3 & &  \\
Spectroscopic &  & & 4950& && 2.0 & 1.65&-1.76$\pm$0.16&-1.72$\pm$0.15 \\
\bottomrule
\end{tabular*}
\end{table*}

\section{Abundance analysis}\label{sec4}
\subsection{Line identification and equivalent width measurements}
For the abundance analysis of HD~1936, we compiled a line list from~\citet{caliskanetal14}~and~\citet{roedereretal14}. In addition, reliability of log~$gf$ values for each line was checked using a NIST database~\citep{NIST_ASD}. The equivalent widths of spectral lines were measured automatically using the IDL routine Binmag~\citep{kochukhove18}. This could be used to fit a Gaussian and rotational profile to the observed lines in the normalized spectrum. For lines stronger than 150~m\AA, we did not used this technique, because Gaussian profile fails to model the Doppler broadened line wing. Our line list is presented in~Table~\ref{line_list}. The CH molecular lines and the hyperfine structure splitting lines of Li, Al, Sc, V, Mn, Co, Cu, Y, Ba, La, and Eu were taken from Kurucz's molecular and hyperfine line lists \footnote{http://kurucz.harvard.edu/linelists}.

\begin{table}
\caption{The line list used in the abundance analysis. The full table is available as supporting online material. A portion of the table is shown here.}
\label{line_list}
\centering
\begin{tabular}{lccc}
\toprule
Atomic Sp. & Wavelength & log $gf$  & Equivalent width\\
           &   (\AA~)     &        & ~m\AA~ \\
 \hline
        Li~I & 6707.00 & 0.002 & syn \\
        O~I & 6300.00 & -9.819 & syn \\
        Na~I & 5682.63 & -0.700 & 10.00 \\
        Na~I & 5688.21 & -0.400 & 18.60 \\
        Mg~I & 4571.10 & -5.691 & 109.10 \\
        Mg~I & 5711.09 & -1.833 & 40.70 \\
        Al~I & 6696.02 & -1.347 & syn \\
        Si~I & 5772.15 & -1.750 & 11.50 \\
\bottomrule
\end{tabular}
\end{table}

\subsection{Determination of chemical abundances}
The sodium, magnesium, silicon, calcium, titanium, chromium, iron, nickel, and zinc abundances were derived from the equivalent width measurements for their unblended atomic lines using the WIDTH9 code~\citep{sbordoneetal04,kurucz05}. We used the SYNTHE spectrum synthesis code~\citep{kurucz93,kurucz05}~for the determination of the lithium, carbon, oxygen, aluminum, scandium, vanadium, manganese, cobalt, copper, and neutron-capture element abundances from their blended or hyperfine structure splitting lines. We adjusted the abundances until the best fit between the synthetic and observed line profiles was achieved. We broadened the synthetic spectra, taking into account the rotational and microturbulent velocities as well as the instrumental profile. 

\section{Results of chemical abundance analysis}\label{sec5}
\begin{figure}
\begin{center}
\hbox{\hspace{-0.2cm}\includegraphics[width=\columnwidth]{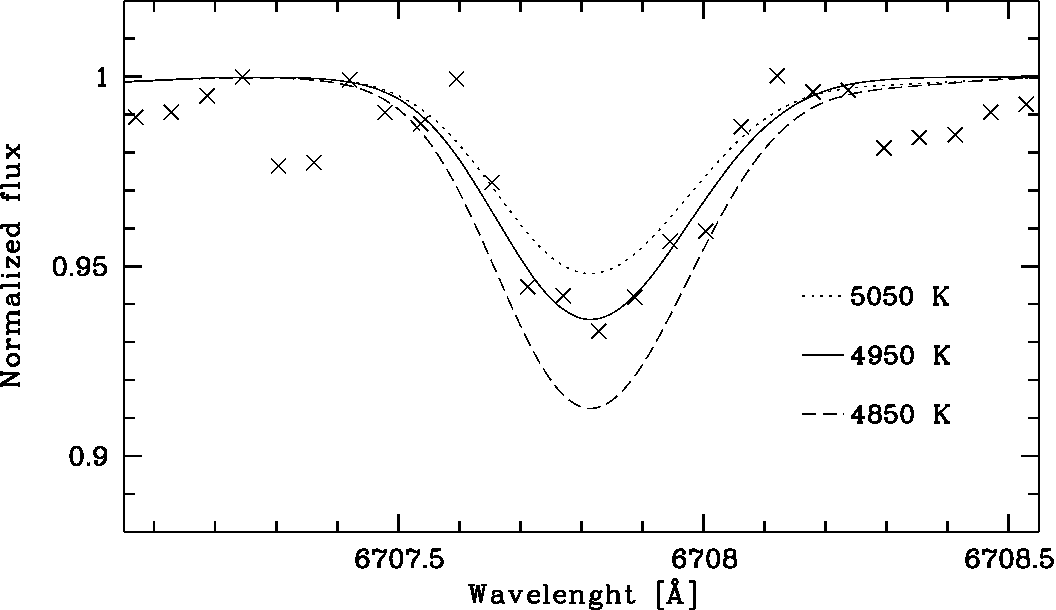}}
\caption{Observed and synthetic lithium lines at 6707~\AA.}
\label{lidoub}
\end{center}
\end{figure}

We found the abundance of lithium in HD~1936 from the 6707~\AA~feature, resulting in a value of A(Li)=1.11. We considered a 3D non-LTE effect of -0.1~dex on the abundance of Li for the adopted atmospheric parameters of the star~\citep{wangetal21}. In that case, our 3D non-LTE lithium abundance is 1.01. If the effective temperature was changed by 100~K, the abundance of Li changed up to 0.15~dex, as shown in Figure~\ref{lidoub}. Therefore, we assumed an uncertainty of 0.15 for the Li abundance. The uncertainties represent the standard deviation of the abundance measurements (log $\epsilon$) of each line, and are listed in~Table~\ref{abutab}. 

\begin{figure}
\begin{center}
\hbox{\hspace{-0.2cm}\includegraphics[width=\columnwidth]{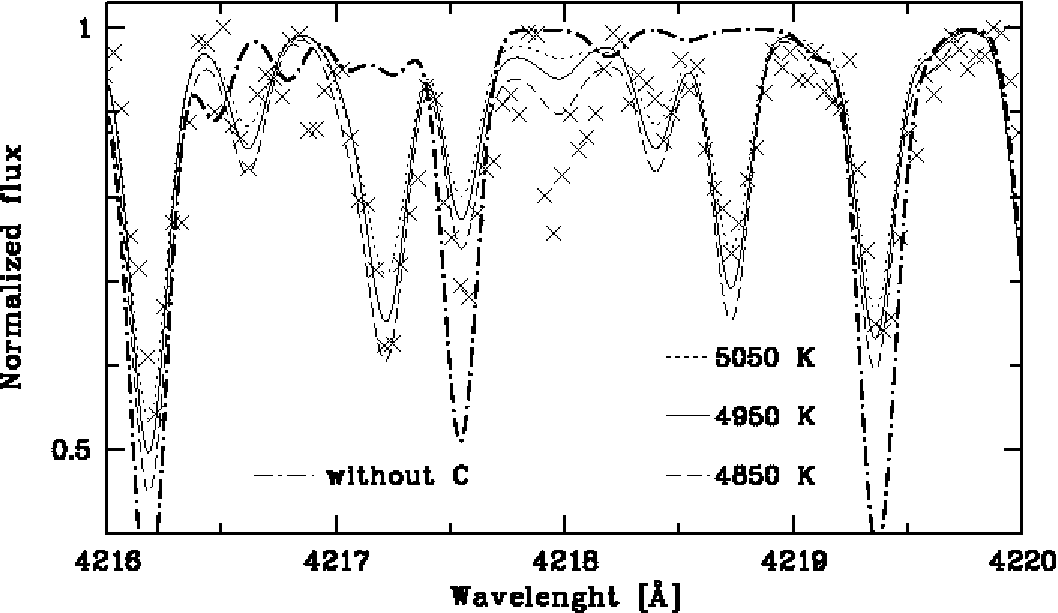}}
\caption{Observed and synthetic CH molecular lines in the region 4220~\AA.}
\label{chband}
\end{center}
\end{figure}

The abundance of carbon was derived from CH molecular lines in the region 4210-4220~\AA, producing a value of A(C)=6.45. The region is given in Figure~\ref{chband}. When the effective temperature was varied by 100~K, we found a change of 0.1~dex in the carbon abundance, which we adopt as the uncertainty this value. The C$_{\rm 2}$ bands were not detectable in the spectra of HD~1936.~\citet{popaetal23} obtained a non-LTE correction of 0.15~dex specifically for the CH molecular feature (G band) in metal-poor stars that have similar parameters with HD~1936. If that value is taken into account, the non-LTE carbon abundance would be 6.60. We measured the oxygen abundance from the [O~I] line at 6300~\AA, which is usually observable in spectra of metal-poor stars. We do not account for the non-LTE effects in the [O~I] 6300~\AA~line, since the line does not suffer from that effect~\citep{amarsietal15}. We adopt an uncertainty of 0.15~dex for oxygen abundance and the other atomic species that have only one detected line in the spectrum, including uncertainties in the atomic data and the equivalent width measurement due to the continuum placement. 

The Na abundance was determined from the 5682~\AA~and 5688~\AA~sodium lines, using their equivalent width measurements. Using the INSPECT webtool~\footnote{http://www.inspect-stars.com/index.php?n=Main.HomePage}, we calculated an average non-LTE correction of -0.07~dex for the LTE sodium abundance from~\citet{lindetal11}.  The final non-LTE sodium abundance was found to be [Na/H]=-2.14. The only detectable aluminum line was the doublet of Al~I at 6696~\AA~in the spectrum of our star. Using the spectrum synthesis method, we obtained an abundance value of [Al/H] as -1.30. The non-LTE abundance correction for the optical aluminum lines was provided by~\citep{lindetal22}. The correction is 0.05~dex when the atmosphere parameters of HD~1936 are considered. Thus, the aluminum abundance changes from -1.30 to -1.25.

We determined all $\alpha$-element (Mg, Ca, Ti) abundances based on the equivalent width measurements of the absorption lines in the spectra of HD~1936. The abundance of magnesium was derived to be [Mg/H]=-1.35 from the 4571~\AA~and 5711~\AA~features. According to~\citet{bergemannetal17}, the 3D non-LTE effects in the lines lead to a reduction in the magnesium abundances from these lines by -0.17~dex and -0.09~dex, respectively. We did not employ the Mg~Ib triplet in the analysis. To find the abundance of Si, we used the three absorption lines, 5772, 5948, 6125~\AA, of silicon. The [Si/H] ratio is -1.21. The non-LTE effects on our Si abundance are negligible~\citep{mashonkinaetal16}. The calcium resonance line 4226~\AA~was not used in the abundance analysis, although it is a prominent feature in the spectrum of HD~1936. We calculated the Ca abundance from the seventeen calcium lines, whose equivalent widths range from 30 to 120~m\AA.~Our [Ca/H] value is -1.47. For the determination of the Ti abundance, we used the absorption features of neutral (11 lines) and once-ionized (8 lines) titanium. Accordingly, the abundance of Ti~I was found to be -1.47 with an uncertainty of 0.17, and [Ti~II/H] was found to be -1.35. The non-LTE effect is about 0.1~dex for the abundances derived from Ti~I lines, and it is close to zero for Ti~II~\citep{mashonkinaetal16}. The difference between [Ti~I/H] and [Ti~II/H] might be due to non-LTE effects in the Ti~I/II lines. Thus, we obtained that the mean $\alpha$-element abundance of HD~1936 is 0.33, indicating the signature of chemical enrichment by Type~II~SNe.

We derived chemical abundances for the iron-peak elements Sc, V, Cr, Mn, Co, and Ni, as well as Cu and Zn. We determined the chromium abundance to be -1.95 from the eight Cr~I lines and -1.70 from the two Cr~II lines. The chromium abundances derived from the Cr~I and Cr~II lines differ from each other by 0.25~dex. A non-LTE abundance correction of 0.3~dex in [Cr~I/H] seems to explain this discrepancy. 
We calculated an average non-LTE correction of 0.31~dex for Cr~I abundance, using the MPIA data base~\footnote{https://nlte.mpia.de/}~\citep{bergemanetal10}. The [Ni/H] value is -1.83 for the star, and [Zn/H]=-1.70. The nickel abundance was calculated from the six Ni~I lines. For the zinc abundance, we used two Zn~I lines at 4722 and 4810~\AA.~We used the synthetic spectrum fitting technique to compute the abundances of Sc, V, Mn, Co, and Cu. The abundance of scandium was measured as -1.80 from the three Sc~II lines. The V and Co abundances were derived from only one absorption line of V~I (at 4881~\AA) and Co~I (at 4749~\AA). The abundances of vanadium and cobalt are -1.83 and -1.70. We calculated a non-LTE correction of 0.35~dex for Co~I abundance, using the MPIA data base~\citep{bergemannetal10, bergemann08}. The manganese abundance of HD~1936 was measured as -2.28 from the six manganese lines at 4754, 4761, 4762, 4765, 4766, and 4823~\AA.~We have calculated the non-LTE corrections ranging from 0.41 to 0.44 for each manganese line~\citep{bergemannetal19, bergemann08}. When applying these corrections to the LTE abundance derived from each line, the determined non-LTE manganese abundance is -1.86.  

\begin{figure}
\begin{center}
\hbox{\hspace{-0.3cm}\includegraphics[width=\columnwidth]{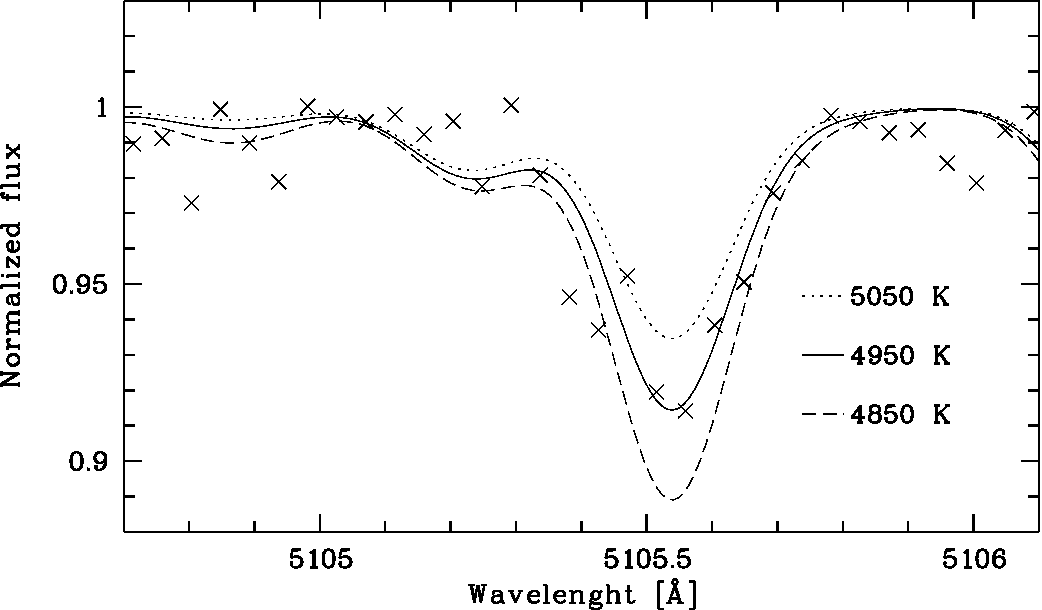}}
\caption{Observed and synthetic Cu~I line at 5105~\AA.}
\label{cu}
\end{center}
\end{figure}

As shown in Fig.\ref{cu}, the copper abundance was found to be A(Cu)=-2.78, derived from the line of 5105~\AA.~The non-LTE effect was considered for the copper abundance. We applied a non-LTE correction of $\sim$0.3~dex, which, according to~\citet{andrievskyetal18}, is appropriate for a star with the parameters of HD~1936. Therefore, the non-LTE A(Cu) was found to be -2.48. Our synthetic spectra were computed by taking into account the hyperfine structure splitting of the Sc~II, V~I, Mn~I, Co~I, and Cu~I components. 

We derived the Sr, Y, Zr, Ba, La, Pr, Nd, and Eu abundances through spectral syntheses. The abundance of strontium was measured at 4607~\AA~from the Sr~I line. We do not use the Sr~II line at 4215~\AA, as it is saturated. The non-LTE correction in Sr~II abundance becomes important as the metallicity decreases, but it is negligible for [Fe/H]>-3. However, the effect is important for Sr~I abundance: it may be up to 0.55~dex for the giants~\citep{hansenetal13}. When we considered a non-LTE correction of 0.5~dex, our non-LTE [Sr/H] abundance was -1.91. For the abundance of Eu, we synthesized the absorption feature at 6645~\AA.~The non-LTE correction for the Eu abundance was given by~\citet{mashonkina14} as $\sim$0.1. The abundance of the light neutron-capture element Y was computed as -2.16, using the Y~II lines at 4398, 4883, 5087, and 5119~\AA.~Two weak Zr~II lines (at 4211~\AA~and 5112~\AA) are detectable in the spectra of HD~1936. We applied a non-LTE correction of 0.18~dex to LTE zirconium abundance, which was computed by~\citet{velichko10}. We extracted the abundance of barium from the 4554, 4934, 5853 and 6141~\AA~lines. \citet{gallagheretal20} expressed that the non-LTE effect for these four lines is small, ranging between -0.05 and -0.1~dex. The abundance of lanthanum derived from the 4804, 5114, and 5290~\AA~lines, that give concordant abundances. The influence of hyperfine structure on the derived abundances of Y, Ba, La, and Eu was taken into account during the synthesis. To calculate the Pr and Nd abundances, we used the two lines of Pr~II (which give concordant abundances) and three lines of Nd~II. Our LTE and non-LTE abundance results and their uncertainties are given in Table~\ref{abutab} and ~\ref{nabutab}, respectively. The solar abundances are taken from~\citet{loddersetal03} and \citet{caffaueta11} for O, C, and Fe. 


\begin{table}
	\centering
	\caption{Derived LTE atmospheric element abundances of HD~1936.}
	\label{abutab}
	\begin{tabular}{lccccl} 
		\hline
		Species &log $\epsilon_{\odot}$& log $\epsilon$ & [X/Fe]& $\sigma_{r}$ & N \\
		\hline
Li~I$^{\rm{syn}}$& 3.28&1.11&-&0.15&1\\
CH$^{\rm{syn}}$ &8.50&6.45&-0.31&0.10&-\\
O~I$^{\rm{syn}}$ &8.76&7.65&0.63&0.15&1\\
Na~I & 6.30 & 4.23 &-0.33 & 0.01 & 2 \\
Mg~I & 7.55 & 6.19 & 0.39 & 0.01 & 2 \\
Al~I$^{\rm{syn}}$& 6.47&5.17& 0.44&0.15&1\\
Si~I & 7.54 & 6.33 &0.53  & 0.11 & 3 \\
Ca~I &6.34 & 4.87 &0.27 & 0.14 & 17\\
Sc~II$^{\rm{syn}}$ & 3.07 & 1.27 &-0.06  & 0.03 & 3 \\
Ti~I &4.92 & 3.45 &0.27  & 0.17& 11\\
Ti~II & 4.92& 3.57 & 0.39& 0.17& 8\\ 
V~I$^{\rm{syn}}$& 4.00 & 2.17 & -0.09 &0.15 & 1\\
Cr~I & 5.65 &3.70 &-0.21 & 0.07 & 8 \\
Cr~II &5.65& 3.95&0.04 & 0.07&2\\
Mn~I$^{\rm{syn}}$ & 5.50 & 3.21 &-0.55 & 0.07 & 6\\
Co~I$^{\rm{syn}}$ & 4.92 & 3.22  & 0.04 & 0.15 & 1\\
Ni~I&6.22 &4.39 &-0.09 & 0.23&6\\
Cu~I$^{\rm{syn}}$ &4.21&1.43&-1.04&0.15&1 \\
Zn~I &4.62&2.92&0.04&0.17&2\\
Sr~I$^{\rm{syn}}$ & 2.91&1.50 &0.33  &0.15& 1  \\ 
Y~II$^{\rm{syn}}$ &2.20 & 0.04 &-0.42  &0.20 &4 \\
Zr~II$^{\rm{syn}}$ &2.60&0.96&0.10&0.01&2\\	
Ba~II$^{\rm{syn}}$ &2.18 &0.24  & -0.21 & 0.15&4  \\
La~II$^{\rm{syn}}$ &1.18&-0.67&-0.11&0.01&3\\
Pr~II$^{\rm{syn}}$ &0.71&-0.63&0.40&0.01&2\\
Nd~II$^{\rm{syn}}$ &1.46&-0.14&0.14&0.08&3\\
Eu~II$^{\rm{syn}}$ &0.51&-0.8&0.43&0.15&1\\
  \hline
 \end{tabular}
 \end{table}

\begin{table}
	\centering
	\caption{The non-LTE atmospheric element abundances of HD~1936.}
	\label{nabutab}
	\begin{tabular}{lccccl} 
		\hline
		Species &log $\epsilon_{\odot}$& log $\epsilon$ & [X/Fe]& $\sigma_{r}$ & N \\
		\hline
Li~I$^{\rm{syn}}$& 3.28&1.01&-&0.15&1\\
CH$^{\rm{syn}}$ &8.50&6.60&-0.16&0.10&-\\
Na~I & 6.30 & 4.16 &-0.40 & 0.01 & 2 \\
Al~I$^{\rm{syn}}$& 6.47&5.22& 0.49&0.15&1\\
Cr~I & 5.65 &4.01 &0.10 & 0.07 & 8 \\
Mn~I$^{\rm{syn}}$ & 5.50 & 3.64 & -0.12 & 0.07 & 6\\
Co~I$^{\rm{syn}}$ & 4.92 & 3.57  & 0.39 & 0.15 & 1\\
Cu~I$^{\rm{syn}}$ &4.21&1.73&-0.74&0.15&1 \\
Sr~I$^{\rm{syn}}$ & 2.91&1.00 &-0.17  &0.15& 1  \\ 
Zr~II$^{\rm{syn}}$ &2.60&0.78 &-0.08 &0.01&2\\	
Ba~II$^{\rm{syn}}$ &2.18 &0.14  &-0.31  & 0.15&4  \\
Eu~II$^{\rm{syn}}$ &0.51& -0.7&0.53 &0.15&1\\
  \hline
 \end{tabular}
 \end{table}
 
\section{Evolutionary status of HD~1936}\label{sec6}
\begin{figure}[t]
\begin{center}
\hbox{\hspace{-0.5cm}\includegraphics[width=\columnwidth]{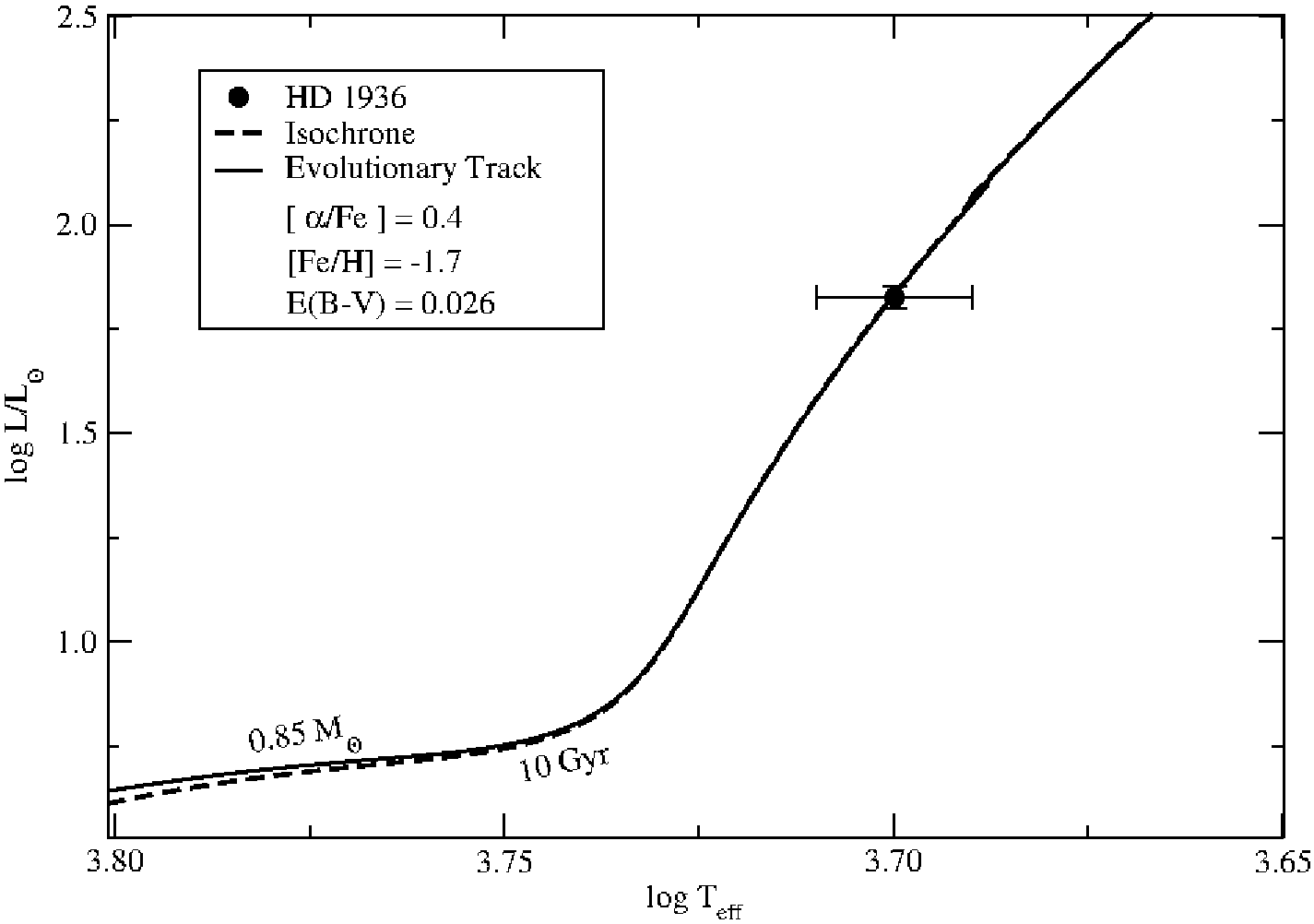}}
\caption{The evolutionary status of HD~1936.}
\label{fig3}
\end{center}
\end{figure}

Using our spectroscopic $T_{\rm eff}$ and the calculated luminosity of HD~1936, we plotted our star on a theoretical HR diagram (Fig.~\ref{fig3}). We calculated the BC$_{\rm V}$ from \citet{alonsoetal99}, the parallax of the star is 3.375~mas from Gaia DR3~\citep{gaiadr3}. The color excess E(B-V) was taken from~\citet{kervellaetal19}. The uncertainties are shown in Fig.~\ref{fig3} with log$T_{\rm eff}$ on the horizontal axis and log$L/L_{\odot}$ on the vertical axis. An evolutionary track with a mass of 0.85$M_{\odot}$~\citep{basti21} and an isochrone at 10~Gyr~\citep{basti21} is consistent with the position of the star on the Hertzsprung-Russell diagram. While our estimated mass of 0.85$M_{\odot}$ for HD~1936 is compatible with that of~\citet{dekaetal18}, it is considerably lower than that of~\citet{kervellaetal19}. We note that the star's position on the HR diagram does not overlap with the evolutionary track corresponding to the mass value of~\citet{kervellaetal19}.

\section{Discussion and Conclusion}\label{sec7}
\begin{figure*}
\begin{center}
\hbox{\hspace{-0.5cm}\includegraphics[width=18cm]{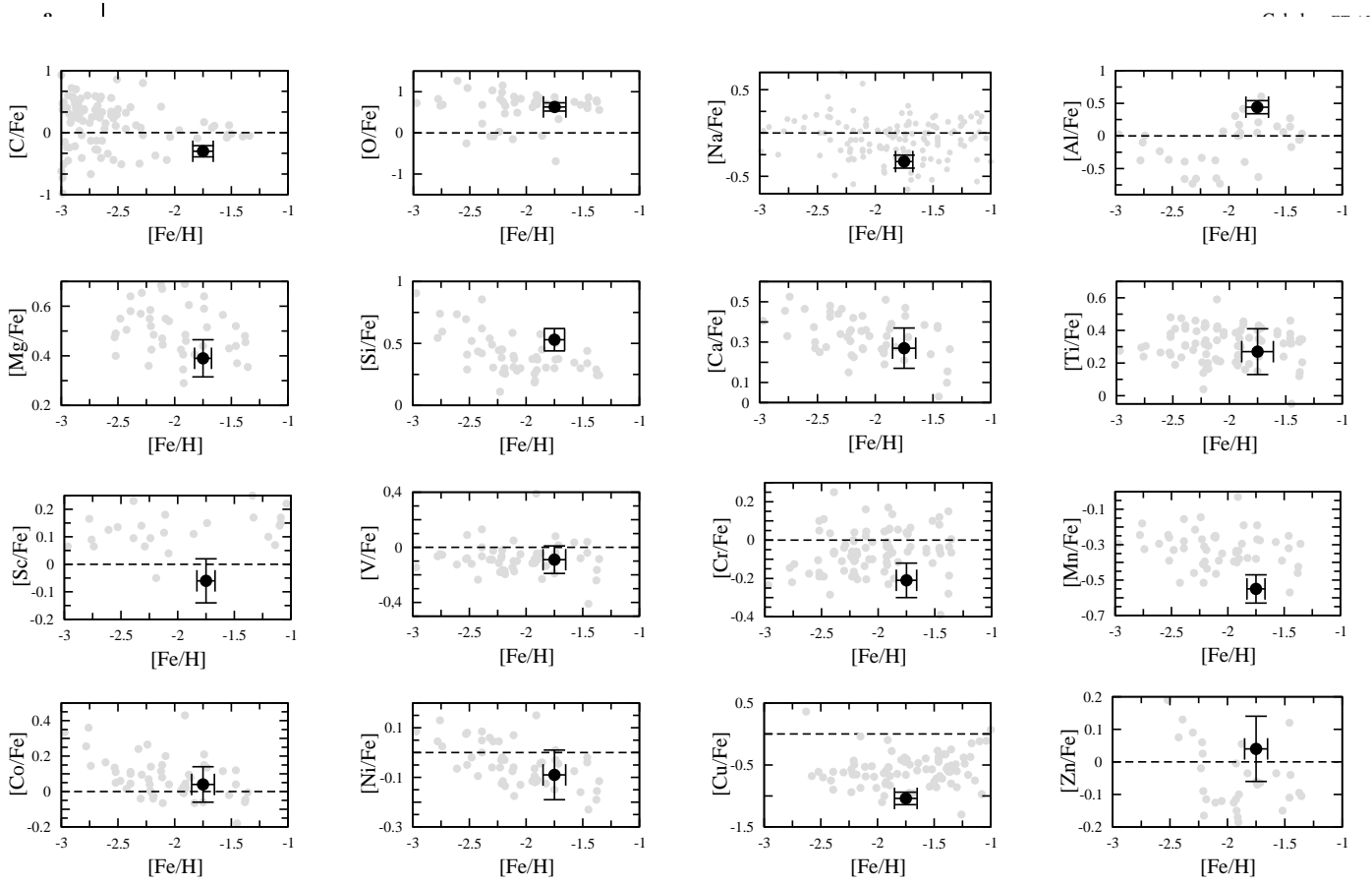}}
\caption{The distribution of the derived LTE abundances in the study as a function of [Fe/H]. The abundance data are taken from~\cite{cescuttietal22, mucciarellietal22, vennetal04, johnson02}. Our star and the field stars are denoted by the black and gray filled circles.}
\label{fig6}
\end{center}
\end{figure*}

The derived abundances of the 29 atomic species are shown as functions of [Fe/H] in Figures~\ref{fig6},~\ref{lit}, and~\ref{neutro}. The light and odd-Z element abundance ratios are mostly compatible with those of the metal-poor field stars. We find that its [Na/Fe] is sub-solar, and [Al/Fe] is super-solar than the solar ratio.

According to stellar evolution models~\citep{iben67}, the abundance of lithium should be low for the giant branch stars. Assuming an initial lithium abundance of 3.3 for the star, it ought not to be higher than 1.5~\citep{charbonnel00} when the star evolves to the red giant branch.~\citet{mucciarellietal22} interpreted that the observed lithium abundance distribution in low-luminosity red giant branch stars shows an agreement with an initial A(Li) close to the cosmological lithium abundance of 2.7~\citep{cyburt08}. A lithium abundance of 1.01$\pm$0.15 seems to be consistent with thin A(Li) plateau of~\citet{mucciarellietal22}, as shown in Fig.\ref{lit}. 

\begin{figure}
\begin{center}
\hbox{\hspace{-0.3cm}\includegraphics[width=\columnwidth]{LinonLTE.eps}}
\caption{The non-LTE A(Li) versus [Fe/H]. The non-LTE lithium abundances of metal-poor giants are taken from~\citep{mucciarellietal22, lietal18, ruchtietal11}. The symbols are as given in Fig.~\ref{fig6}.}
\label{lit}
\end{center}
\end{figure}

HD~1936 is not enhanced in carbon, as is similar to thin A(Li) plateau giants, and it has a [C/Fe] value of -0.31. According to the classification proposed by~\citet{bonifacioeta18}, it should be a carbon-normal star, with [Fe/H]>-4 and [C/Fe]<1. The derived oxygen abundance ratio of HD~1936 shows an agreement with those of the field giants with similar metallicity. 

\begin{figure}
\begin{center}
\hbox{\hspace{-0.5cm}\includegraphics[width=\columnwidth]{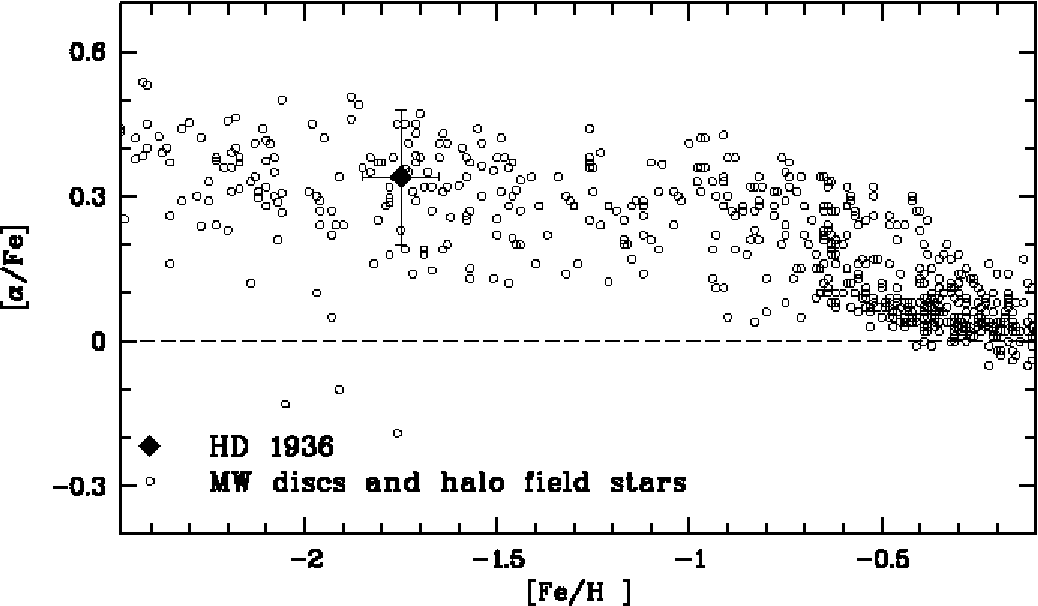}}
\caption{[$\alpha$/Fe] versus [Fe/H]. The references are as given in Fig.~\ref{fig6}.}
\label{alpha}
\end{center}
\end{figure}

HD~1936 is $\alpha$-enhanced with the [(Mg, Ca, Ti)/Fe]$\sim$0.33. This abundance ratio is consistent with the distribution of $\alpha$-element abundances of a classical metal-poor star in the Galactic halo, as given in Fig.\ref{alpha}. This value suggests that the contribution from Type~II~SNe dominates in the star's chemical inventory, as expected from metal-poor halo stars.

Most of the iron-peak element abundance ratios display nearly solar ratios, [Sc, V, Cr, Mn, Ni, Zn/Fe]$\sim$0, it is unsurprising for stars with similar metallicity. This may be a signature that these iron-peak elements share the same production channels as iron. We find a super-solar cobalt abundance with a value of 0.39, if we consider the non-LTE effect on its abundance. This is unusual at a metallicity of -1.74, because the observational trend for LTE cobalt abundance is typically an increase with decreasing metallicity. Our LTE cobalt abundance, [Co/Fe]=0.04, is in agreement with the observational trend. The copper exhibits a sub-solar abundance ratio, we will discuss this further.

The abundance patterns of neutron-capture elements are shown in Fig.\ref{neutro}. Our abundances exhibit a similar trend in [Sr, Y, Zr, Ba, La, Nd, Eu/Fe] and [Fe/H] as compared to the metal-poor halo stars. 

\begin{figure*}
\begin{center}
\hbox{\hspace{1.0cm}\includegraphics[width=15cm]{abundancesSrtoEu.eps}}
\caption{The LTE abundance ratios [Sr, Y, Zr, Ba, La, Nd, Eu/Fe] versus [Fe/H] for HD~1936. The symbols are as given in Fig.~\ref{fig6}. The abundance data are taken from~\citep{sarafetal23,mcwilliametal95,ryanetal96,burrisetal00,fulbright00,westinetal00,cowanetal02,cayreletal04, 2004ApJ...603..708C,2004ApJ...607..474H, 2005ApJ...632..611A, 2005A&A...439..129B, 2006ApJ...645..613I, 2006AJ....132.1714P,2008ApJ...681.1524L}.\citep{2009A&A...504..511H, 2010ApJ...723L.201A, 2010ApJ...724..975R, 2013ApJ...778...56C, 2013ApJ...775L..27J, 2012A&A...545A..31H, 2013ApJ...771...67I,2014A&A...569A..43M, 2014ApJ...784..158R, 2015ApJ...807..171J, 2015Natur.527..484H,2015RAA....15.1264L}. \citep{2015ApJ...807..173H, 2016MNRAS.460..884H, 2017ApJ...844...18P,2018ApJ...864...43C,2018ARNPS..68..237F, 2018ApJ...858...92H, 2018MNRAS.481.1028H, 2018ApJ...865..129R, 2018ApJ...868..110S, 2019ApJ...875...89M,2019A&A...627A.173V,2019NatAs...3..631X}. \citep{2020ApJ...898...40C,2020ApJS..249...30H,2020ApJ...897...78P, 2020ApJ...905...20R}.}
\label{neutro}
\end{center}
\end{figure*}

\begin{figure}
\begin{center}
\hbox{\hspace{-0.5cm}\includegraphics[width=\columnwidth]{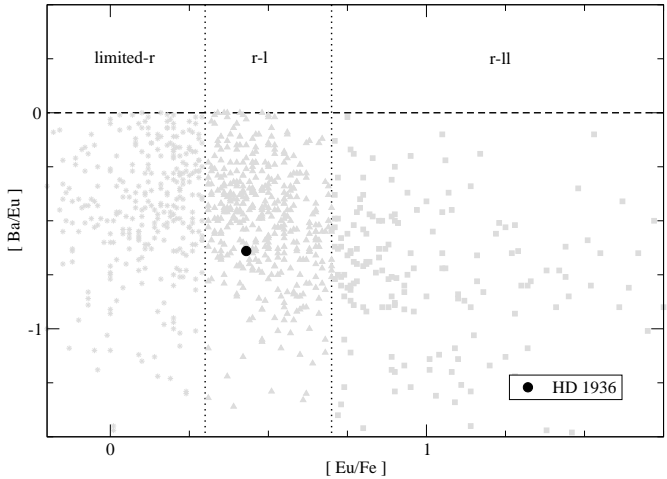}}
\caption{[Ba/Eu] versus [Eu/Fe] for HD~1936. The references are as given in Fig.~\ref{neutro}.}
\label{baeu}
\end{center}
\end{figure}

We investigate whether the star is enriched in neutron-capture elements based on the abundance ratios of [Ba/Fe], [Eu/Fe], and [Ba/Eu]. The barium is dominantly produced by the slow neutron-capture process (s-process), while most of europium comes from the rapid neutron-capture process (r-process). Its [Ba/Fe] is -0.21, its [Eu/Fe] is 0.43, and its [Ba/Eu] is -0.64. According to the definition of sub-classes of metal-poor stars given by~\citep{beers05}, HD~1936 is a moderately r-process enhanced (r-I) star. Among the known metal-poor stars, 1346 r-I stars have been discovered so far~\citep{shanketal23}. As shown in Fig.\ref{baeu}(Fig.11 of~\citet{sarafetal23}, the star falls within the region of the r-I sub-class, rather than the other r-process enhanced star sub-classes limited-r~\citep{arconesthielemann23} and highly r-process enhanced (r-II). The Pr and Nd abundances exhibit an overabundance with respect to iron, in common with the other r-I stars~\citep{siqueetal14}. 

\begin{figure}
\begin{center}
\hbox{\hspace{-0.5cm}\includegraphics[width=\columnwidth]{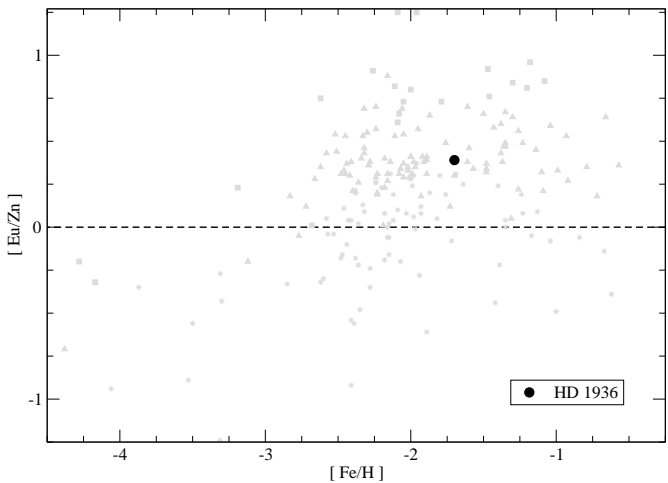}}
\caption{[Eu/Zn] versus [Fe/H] for HD~1936. The references are as given in Fig.~\ref{neutro}.}
\label{euzn}
\end{center}
\end{figure}

The [Eu/Zn] vs [Fe/H] plot shown in Fig.~15 of~\citet{sarafetal23} is interesting. When we put the star on this plot, we notice that its position is well-matched with the trend observed in their r-I star. As shown in Fig.~\ref{euzn}, the [Eu/Zn] has an increase trend with increasing metallicity. There are several sites and channels for the zinc production, even though it is produced mainly during explosive Si-burning. According to~\citet{kobayashietal06}, the [Zn/Fe]$\sim$~0 ratio for the stars at metallicity regime of HD~1936 can be explained only by the large contribution of hypernovae. Furthermore,~\citet{yongetal21} states that a hypernova with progenitor mass of 25$M_{\odot}$ could produce both the r-process elements and  iron-peak elements during explosive nuclear burning. As is known, the major astrophysical sites of r-process-enrichment still remain uncertain, but we do have some predictions about what these sites could potentially be, such as neutrino-driven winds in Type~II~SNe, neutron star mergers~\citet{arconesthielemann23, watsonetal19}. Thus,
a high [Eu/Zn] ratio of HD~1936 may be indicated that its chemical inventory has a contribution from neutron star merger events.  

The star's [Cu/Fe] of -0.74, combined with the metallicity of -1.74, may indicate that the star is one possible tracer of the rare descendants of the very massive first-generation stars~\citep{salvadorietal19}. Furthermore, the [Mg/C] ratio of the star provides insights into whether the gas in which the star was formed was polluted by only one SN event or by more than one~\citep{hartwigetal18}. We found a [Mg/C] ratio of 0.69 for HD~1936, indicating enrichment due to more than one SN. 





The results of~\citet{shapiroetal23} suggest that planets orbiting stars with a low metallicity offer the most promising prospects for finding terrestrial life of a sophisticated nature. According to~\citet{kervellaetal19}, HD~1936 has a substellar companion with a mass of 18.35$M_{J}$. If so, a companion with such a mass must be a giant planet, not a rocky one.~\citet{reggianimelendez18} has recently provided spectroscopic evidence of planetary formation at [Fe/H]=-1.4, while the occurrence of massive planet formation around such stars remains largely uncertain. Host stars with giant planets tend to be more metal-rich than those with smaller planets~\citep{ghezzietal10}. Precise radial velocity measurements are necessary to determine whether HD~1936 indeed hosts a massive planet. A planet-hosting giant star with [Fe/H]=-1.74 would be interesting for research into planet formation scenarios.

\bibliography{Wiley-ASNA}

\section*{Acknowledgments}
The data in this study were obtained with the T80 telescope at the Ankara University Astronomy and Space Sciences Research and Application Center (Kreiken Observatory) with project number of 23A.T80.04. The authors thank an anonymous referee for his/her useful comments in improving the article. This work has also made use of the SIMBAD database operated at the CDS, Strasbourg, France, and of NASA’s Astrophysics Data System.
\end{document}